\documentclass[12pt,draft]{amsart}
\usepackage{amsmath,amsthm,amscd,amssymb}
\usepackage{latexsym}

\newcommand{\curl}{\operatorname{curl}}
\newcommand{\rsp}{\operatorname{rsp}}
\newcommand{\ess}{\operatorname{ess}}
\newcommand{\dv}{\operatorname{div}}
\newcommand{\wlim}{\operatorname{w-lim}}
\newcommand{\vel}{\operatorname{vel}}

\newcommand{\dom}{\operatorname{Dom}}

\theoremstyle{plain}
\newtheorem{THEOREM}{Theorem}
\newtheorem{COROL}[THEOREM]{Corollary}
\newtheorem{LEMMA}[THEOREM]{Lemma}
\newtheorem{PROP}[THEOREM]{Proposition}
\newtheorem{REMARK}[THEOREM]{Remark}

\theoremstyle{definition}

\theoremstyle{remark}

\newcommand{\R}{\ensuremath{\mathbb{R}}}   
\newcommand{\C}{\ensuremath{\mathbb{C}}}   
\newcommand{\T}{\ensuremath{\mathbb{T}}}   

\def \a {\alpha}
\def \b {\beta}
\def \d {\delta}
\def \g {\gamma}

\def \f {\varphi}

\def \l {\lambda}

\def \n {\nabla}
\def \s {\sigma}

\def \t {\tau}
\def \o {\omega}

\def \< {\langle}
\def \> {\rangle}
\def \ra {\rightarrow}
\def \ss {\subset}

\DeclareMathOperator{\diver}{div} %
\DeclareMathOperator{\re}{Re} %
\newcommand{\rest}[2]{#1\raisebox{-0.3ex}{\mbox{$\mid_{#2}$}}}

\newcommand{\kc}[1]{[D\f_t(#1)]^{-\top}}

\begin{document}

\title[Euler equation and Lyapunov exponents]{Essential spectrum of
the linearized 2D Euler equation and Lyapunov-Oseledets
exponents}
\author{Roman Shvydkoy and Yuri Latushkin}
\address{Department of Mathematics \\ University of Missouri\\
Columbia, MO 65211} %
\email{shvidkoy@math.missouri.edu, yuri@math.missouri.edu}%
\thanks{The authors wish to thank S. Friedlander
and M. Vishik for many stimulating discussions.
Second author was partially
supported by the Twinning Program of the
 National Academy of Sciences
and National Science Foundation, and by the
 Research Council and Research Board
of the University of Missouri.}%
\date{\today}%

\begin{abstract}  The linear stability
of a steady state solution of 2D Euler equations of an ideal fluid
is being studied.
 We give an explicit geometric construction of approximate
eigenfunctions for the linearized Euler operator $L$ in vorticity
form acting on Sobolev spaces on two dimensional torus. We show
that each nonzero Lyapunov-Oseledets exponent for the flow induced
by the steady state contributes a vertical line to the essential
spectrum of $L$. Also, we compute the spectral and growth bounds
for the group generated by $L$ via the maximal Lyapunov-Oseledets
exponent. When the flow has arbitrarily long orbits, we show that
the essential spectrum of $L$ on $L_2$ is the imaginary
axis.\end{abstract}

\maketitle

\section{Introduction}

Let $u=u(x)$ be a $C^\infty$-steady state solution of the Euler
equations governing the motion of an inviscid ideal fluid:
\begin{equation}
\partial_tu+\langle u,\nabla \rangle u+\nabla P=0,\quad \dv u=0.
\label{NLEE}\end{equation}
Here $u$ is the velocity, $P$ is the pressure, and $\langle \cdot
,\cdot \rangle$ denotes the scalar product. The study of the
spectrum of the linearized Euler operator $L$ obtained by
linearization of the Euler equations about the steady state and
the spectrum of the group $\{e^{tL}\}$ has a long history, see
\cite{C,DH,DR,FH,Lin,Y}.

Recently, an important breakthrough has been made in understanding
the {\em essential spectrum} of $L$ and $e^{tL}$, see
\cite{FV91a,FV92,FSV97,FSV99,LH1,LH2,V96,FV93a} and the
bibliography therein. In particular, using asymptotic expansions
for integral Fourier operators, the boundary of the essential
spectrum of $e^{tL}$ (in dimensions two and three) was related to
the maximal Lyapunov exponent of a so-called {\em bicharacteristic
amplitude system}, see \cite{V96,FV93a,S2} and also \eqref{BAS}
below. These equations are obtained by substituting a high
frequency oscillating anzats into Euler equations. As a result, it
was discovered in \cite{FV91a,FV92} that the presence of positive
Lyapunov exponents for the flow induced by the steady state leads
to the linear hydrodynamic
 instability of the fluid. Later,
using the bicharacteristic amplitude system, results from
\cite{V96}, and a construction of highly oscillating approximate
eigenfunctions for $L$, the boundaries of the essential spectra of
$L$ and $e^{tL}$ for velocity in $L_2$ were related in
\cite{LV02}. Note that in dimension two the maximal Lyapunov
exponent of the bicharacteristic amplitude system is equal
to the maximal Lyapunov exponent of the flow induced by the steady state.

In the current paper (for dimension two) we propose an approach
that does not require either the use of the bicharacteristic
amplitude system or the high frequency asymptotic expansions, and
give an explicit construction of approximate eigenfunctions for
the linearized Euler operator on all Sobolev spaces working
directly with the flow induced by the steady state. This
construction is related to the one used in \cite{LV02}. Also, we
take a look inside the essential spectrum and show that each
nonzero Lyapunov-Oseledets exponent of the flow contributes a
whole vertical line to the spectrum. This also gives a formula for
the boundaries of the essential spectra of $L$ and $e^{tL}$ in
terms of the Lyapunov exponents for the flow generated by $u$.

In the subsequent work \cite{SvLat}, using the results of the current
paper, we proved that the essential spectrum of $L$ in dimension two
fills a solid vertical strip. However, the formulas for the approximate
eigenfunctions presented here allow one to prove that the Lyapunov-Oseledets
exponents generate vertical lines in the essential spectrum of the linearized
surface quasi-geostrophic equation \cite{Sv}.

We study the linearized Euler operator $L$ in vorticity form,
\begin{equation}\label{linvort}
Lw=-\langle u,\nabla \rangle w-\langle \curl^{-1}w,
\nabla \rangle \curl u,
\end{equation} on the Sobolev spaces $H^0_m=H^0
_m(\mathbb{T}^2;\mathbb{C})$, $m\in\mathbb{Z}$, of scalar functions $w$
having zero means $\int wdx=0$ on the $2$-torus
$\mathbb{T}^2=\mathbb{R}^2/2\pi\mathbb{Z}^2$. We set
$H^0_0=L^0_2(\mathbb{T}^2;\mathbb{C})$. See Section \ref{secnp} for an explanation
how the operator $L$ in vorticity form is related to the linearization
of the Euler equation \eqref{NLEE} for velocity.

Our main observation is that in the representation $L=-A+K$, where
$Kw=-\langle \curl ^{-1}w,\nabla\rangle \curl u$ is a compact
operator, the operator $A$, $Aw=\langle u,\nabla\rangle w$,
generates a so-called {\em evolution}, or {\em Mather semigroup}
$e^{tA}w=w\circ \varphi_t$.  Here and below $\varphi_t :x_0
\mapsto x(t;x_0)$ is the flow on $\mathbb{T}^2$ induced by the
steady state velocity field, that is, by the solutions of the
equation $\partial_t{x}(t)=u(x(t))$. Note that $A^*=-A$ with
respect to the $L_2$-paring.

The spectral theory of the evolution semigroups is fairly well
understood, see \cite{CL} and the bibliography therein. In
particular, there are several known ways to construct approximate
eigenfunctions for the operators $A$ and $e^{tA}$. We stress that
the construction of approximate eigenfunctions proposed in the
current paper is much easier than those in \cite{CL}.

\section{Notation and Preliminaries}\label{secnp}

For an operator $B$ on a Hilbert
space $\mathcal{H}$ we denote by $\sigma (B)=\sigma (B; \mathcal{H})$ its spectrum, that is, the
set of all $z\in\mathbb{C}$ such that $B-zI$ does not have a bounded inverse.
We denote by $\sigma _{\ess}(B)= \sigma _{\ess}(B;\mathcal{H})$ the essential
(Weyl) spectrum, that is, the set of all $z\in\sigma(B)$ such that $z$ is not
an isolated eigenvalue of finite algebraic multiplicity (see, e.g., \cite{EE}
for a detailed discussion of various notions of the essential spectrum). We
let $\rsp(B)$ and $\rsp\ess(B)$ denote the spectral radius and essential
spectral radius of a bounded operator $B$. Recall,
that Nussbaum's formula
for essential spectral radius reads \cite{Nusb}:
\begin{equation}\label{nusbform}
 \rsp\ess(B)=\lim_{n\to\infty}\Big(\inf_K\|B^n+K\|\Big)^{1/n},
\end{equation}
where the $\inf$ is taken over the set of compact operators on $\mathcal{H}$.

If $B$ is a generator of a strongly
continuous semigroup $\{e^{tB}\}_{t\ge0}$ on $\mathcal{H}$, then
\[\omega(B)=t^{-1}\log \rsp(e^{tB})=
\lim_{\tau\to\infty}\tau^{-1}\log\|e^{\tau B}\|\]
and $\omega_{\ess}(B)=t^{-1}\log
\rsp\ess (e^{tB})$, $t\neq 0$, denote the growth bound and the
essential growth bound of the semigroup. Let \[s(B)=\sup
\{ \re z:z\in \sigma (B)\}\] denote the spectral bound. Remark that $\omega(B)\ge
s(B)$ for all strongly continuous semigroups on $\mathcal{H}$. However, the
inverse inequality is, generally, {\em false}, see, e.g., \cite{EN} for a
discussion and further references on this topic.

We say that $z\in \mathbb{C}$ is an {\em approximate eigenvalue} and a
sequence $\{ g_n\}_{n=1}^\infty$ is an {\em approximate
eigenfunction} for $B$ if $\|g_n\|=1$, $g_n\in \dom B$,
 and $\lim_{n\to \infty}\| (B-z)g_n\|=0$. We say that an
approximate eigenfunction is {\em weakly null} if, in addition,
$\wlim_{n\to \infty}g_n=0$ for the weak limit.

For $m\in\mathbb{N}$ we let $\mathcal{B}_m$ denote the set of
$m$-linear operators $B$ with
$\|B\|_{\mathcal{B}_m}=\sup\{|B(v_1,\ldots,v_m)|:
|v_1|=\ldots=|v_m|=1\}$. We write $c$ for a generic constant,
$a\lesssim b$ if $a\le cb$, and denote by $\mathbf{1}_{[a,b]}$ the
characteristic function of the interval $[a,b]$. We use symbol
``$\top$" to denote transposition.

Let $\mathcal{M}=\mathcal{M}(\{\varphi_t\})$ denote the set of
$\varphi_t$-invariant Borel probability measures on
$\mathbb{T}^2$, and $\Sigma$ denote the set of all Lyapunov-Oseledets
exponents $\lambda=\lambda(\nu)$ for the differential $\{
D\varphi_t\}_{t\in \mathbb{R}}$ given for each $\nu \in \mathcal{M}$
by the Oseledets' Multiplicative Ergodic Theorem \cite{Os}, see Section
\ref{prfs}
for more details. Denote
$\Lambda =\sup \{ \lambda(\nu) :\nu \in \mathcal{M}\}$.
 Recall, see
\cite[Thm.8.15]{CL}, that
\begin{equation}\label{LyapF}
\Lambda=\lim_{t\to\infty}
t^{-1}\log\max_{x\in\mathbb{T}^2}\|D\varphi_t(x)\|.
\end{equation}

Going back to Euler equation \eqref{NLEE}, consider its linearization
about the steady-state $u$. The corresponding linear operator $L_{\vel}$
acts on (divergence free, velocity) vector fields by the rule
\begin{equation}\label{velop}
L_{\vel}v=-\langle u, \nabla\rangle v-\langle v,\nabla\rangle
u-\nabla P.\end{equation}
The operator $L_{\vel}$ with the maximal domain $\{v\in H_m^s: L_{\vel}v\in
H_m^s\}$ will be considered on the space
$H_m^s=H_m^s(\mathbb{T}^2;\mathbb{C}^2)$ of divergence free vector fields
from the Sobolev space $H_m(\mathbb{T}^2;\mathbb{C}^2)$, $m\in\mathbb{Z}$.
Note that because we
are in the two dimensional situation, vorticity $w=\curl v$ is a {\em scalar}
function. Throughout, by $\curl v$ we mean the scalar $\curl$ of a two
dimensional vector field $v=(v_1, v_2)^\top$, that this, $w=\curl
v=-\partial_2v_1+\partial_1v_2$. If $w$ is a scalar function on $\mathbb{T}^2$
having zero mean, then we denote by $v=\curl^{-1}w$ the unique solution of the
system $\curl v=w$, $\diver v=0$ on $\mathbb{T}^2$.

Passing to the Fourier transform
$w(x)=\sum_{\mathbf{k}\in\mathbb{Z}^2}w_{\mathbf{k}}e^{i\mathbf{k}\cdot x}$,
$w_0=0$, $\mathbf{k}=(k_1,k_2)^\top\in\mathbb{Z}^2$, $x\in\mathbb{T}^2$,
we have:
\begin{equation}\label{formcurl}
v(x)=\curl^{-1}w(x)=\sum_{\mathbf{k}\in\mathbb{Z}^2\setminus\{0\}}
\dfrac{(-k_2, k_1)^\top}{\|\mathbf{k}\|^2}w_\mathbf{k}e^{i\mathbf{k}\cdot
x}.\end{equation}
Therefore, the operator $\curl: H_m^s(\mathbb{T}^2;\mathbb{C}^2)\to
H_{m-1}^0(\mathbb{T}^2;\mathbb{C})$, $m\in\mathbb{Z}$, is an isomorphism.

For the operator $L_{\vel}$ defined in \eqref{velop}, and the operator
$L$ defined in \eqref{linvort}, we
note the identity
\begin{equation}\label{velvort}
\curl^{-1} L \curl v= L_{\vel} v.
\end{equation}
Indeed, since both $u$ and $v$ in \eqref{velop} are divergence free, by
standard vector identities we infer:
\begin{eqnarray*}
\curl L_{\vel} v &=& \curl\Big(-\langle u, \nabla\rangle v-
\langle v,\nabla\rangle
u-\nabla P\Big) \\
&=& - \langle u, \nabla\rangle\curl v - \langle v,\nabla\rangle \curl u
=L \curl v.\end{eqnarray*}
\begin{REMARK}\label{comprem}
{\em Since the operators $L:H_m^0(\mathbb{T}^2;\mathbb{C})\to
H_m^0(\mathbb{T}^2;\mathbb{C})$ and $L_{\vel}:
H_{m-1}^s(\mathbb{T}^2;\mathbb{C}^2)\to H_{m-1}^s(\mathbb{T}^2;\mathbb{C}^2)$
are similar by \eqref{velvort}, we conclude that their spectra in respective
spaces are equal.}\hfill$\Diamond$\end{REMARK}

Finally, since the symbol
$\{(-k_2,k_1)^\top/\|\mathbf{k}\|^2\}_{\mathbf{k}\in\mathbb{Z}^2}$
in \eqref{formcurl} tends to zero as $\|\mathbf{k}\|\to\infty$, we remark
that the operator $K:H_m^0\to H_m^0$, acting on the space of
{\em $2\pi$-periodic} functions by the rule
$Kw=-\langle \curl ^{-1}w,\nabla\rangle \curl u$,
is compact for all $m\in\mathbb{Z}$.

\section{Main Results}\label{mres}

Let $p(x)=\inf \{ t>0:\varphi_tx=x\}$ denote the prime period of
$x\in \mathbb{T}^2$. We set $p(x)=\infty$ if the point $x$ is
nonperiodic. We say that the flow $\{ \varphi _t\}_{t\in
\mathbb{R}}$ has {\em arbitrarily long trajectories} if for each
$N\in \mathbb{N}$ there is an $x\in \mathbb{T}^2$ such that
$p(x)\geq N$.

\begin{THEOREM}\label{T:main} If $m\in \mathbb{Z}\backslash \{0\}$
then $m\Sigma \backslash \{0\} +i\mathbb{R}\subset
\sigma_{\ess}(L;H^0_m)$.\end{THEOREM}

\begin{THEOREM}\label{T:main2} If $\{\varphi_t\}$ has arbitrarily
long trajectories, then
$$\text{\bf (a)}\quad
\sigma_{\ess}(L;L^0_2)=i\mathbb{R}\quad\text{and}\quad \text{\bf
(b)}\quad i\mathbb{R}\subset \sigma_{\ess} (L;H^0_m), \quad m\in
\mathbb{Z}.$$\end{THEOREM}

These theorems also hold if $\sigma_{\ess}(L)$ is replaced by
$\sigma(A)$.

Note that the assumption on $\{ \varphi_t\}$ in
Theorem~\ref{T:main2} is essential. Indeed, if $u=(1,0)^\top$
then $\sigma (L;L^0_2)=2\pi i\mathbb{Z}$.
The assumption in Theorem~\ref{T:main2} holds for many
flows on $\mathbb{T}^2$ as shown, e. g., in
the following proposition
proved in Appendix.

\begin{PROP}\label{THM4.1} If $u$ is not
identically zero and has at least two distinct stagnation points,
then $\{ \varphi_t\}$ has arbitrarily long orbits.\end{PROP}

Theorems \ref{T:main}--\ref{T:main2} and the spectral inclusion
$\exp(t\sigma(L))\subset\sigma(e^{tL})$, see \cite[Thm.IV3.6]{EN},
 show that each nonzero
$\lambda\in\Sigma$ generates a circle in $\sigma_{\ess}(e^{tL})$.
The bounds of $\sigma_{\ess}(e^{tL})$ are given in the next
result.

\begin{THEOREM}\label{FESOM} If $m\in \mathbb{Z}$
then $\omega_{\ess}(L)=\omega_{\ess}(-L)=|m|\Lambda$ on
$H^0_m$.\end{THEOREM}

The proofs of these results are given in the next section. Passing
to the dual space $H^0_{-m}$, if necessary, we can assume that
$m\geq 0$. Our plan for the proof of Theorems~\ref{T:main} and
\ref{T:main2} is to construct a sequence of approximate
eigenfunctions for the operator $A$, whose supports are stretched
along a suitably chosen streamline and tend to zero in measure.
That is, for Theorem~\ref{T:main}, we will construct for each
$\lambda \in \Sigma\backslash \{0\}$ and $\xi \in \mathbb{R}$ a
weakly null approximate eigenfunction $\{g_n\}\subset H^0_m$ for
the approximate eigenvalue $\lambda +i\xi$ for $A$. Then we will
use the compactness of $K$ to extract a subsequence on which $\|
Kg_j\|\to 0$. Since $L=-A+K$, this way we will produce an
approximative eigenfunction for $L$.
The proof of Theorem~\ref{FESOM} follows from the equality
$\omega(A)=\omega_{\ess}(A)=m\Lambda$ proved in Section \ref{prfs},
compactness of $K$,
and Nussbaum's formula \eqref{nusbform} for the essential spectral radius
that implies $\omega_{\text{ess}}(L)=\omega_{\text{ess}}(-A)$.
\begin{COROL}\label{SEQOM}
If $m\in \mathbb{Z}$, then $s(L)=\omega(L)$ on $H^0_m$.
\end{COROL} This holds since $\omega (L)=\max
\{s(L),\omega_{\ess}(L)\}$ by \cite[Cor.IV.2.11]{EN}, and
 $s(L)\geq |m|\Lambda =\omega_{\ess}(L)$
by Theorem~\ref{T:main}, Theorem~\ref{FESOM}, and identity
$L(\curl u)=0$ (for $m=-1$ see \cite{LV02} for a different proof).

Let $L_{\vel}v=-\langle u, \nabla\rangle v-\langle v,\nabla\rangle
u-\nabla P$ be the linearized Euler operator \eqref{velop} in velocity form
acting on the space $H^s_m=H^s_m(\mathbb{T}^2;\mathbb{C}^2)$ of
divergence free vector fields $v$. We have $\sigma
(L;H^0_m)=\sigma (L_{\vel};H^s_{m+1})$, $m\in \mathbb{Z}$, by Remark
\ref{comprem}. Thus,
all results above can be reformulated for $L_{\vel}$. In
particular, $\omega_{\ess}(L_{\vel};H^s_m)=|m-1|\Lambda$ and
$\omega (L_{\vel};H^s_m)=s(L_{\vel};H^s_m)$, $m\in \mathbb{Z}$. To
relate these results to the work in \cite{V96,FV93a}, let
\begin{equation*}\begin{split} \mu=\lim_{t\to
+\infty}t^{-1}\log \max &\{
|b(t;x_0,\xi_0,b_0)|:(x_0,b_0,\xi_0)\in \mathbb{T}^2\times
(\mathbb{R}^2)^*\times\mathbb{R}^2,\\
&|b_0|=|\xi_0|=1\quad \text{and} \quad \xi_0\bot
b_0\}\end{split}\end{equation*} denote the maximal Lyapunov
exponent for the $b$-equation of the following {\em
bicharacteristic amplitude system}:
\begin{equation}\label{BAS} \begin{split}\partial_t{x} &= u(x),
\quad \partial_t {\xi}=-(Du)^\top\xi,\\
\partial_t{b}& = -(Du)b+2\langle (Du)b,\xi\rangle \xi
|\xi|^{-2}.\end{split}\end{equation} Here and below $Du=\partial u/\partial
x$ is the Jacobi matrix, and we
write $Du^{-\top}=((Du)^\top)^{-1}$. Note that $|b(t)||\xi (t)|$
is a first integral for \eqref{BAS}, see \cite{FV92}. The
Multiplicative Ergodic Theorem, therefore, implies that $\Sigma$
coincides with the Lyapunov-Oseledets spectrum of the cocycle
generated by $b$\,-\,equation in \eqref{BAS} and, in particular,
$\Lambda =\mu$, cf. \cite{FV92}. It was proved in \cite{V96,FV93a}
that $\omega_{\ess}(L_{\vel};L^s_2)=\mu$. Thus, we have a
generalization of this formula for any $m\in \mathbb{Z}$. Also,
note the estimate $\omega_{\ess}(L_{\vel},H^s_m)\geq \mu_m$ given
in \cite{FV91a}. Here $\mu_m$ is the maximal Lyapunov exponent for
$(1+|\xi|^2)^{m/2}b(t)$. This estimate is in tune with the
inequality ``$\geq$" in Theorem~\ref{FESOM}.
\begin{REMARK}{\em In the subsequent paper \cite{SvLat}, using the results above
and a modification of the construction used in the proof of
Theorems~\ref{T:main} and \ref{T:main2}, we show that, in fact,
$\sigma_{\ess}(L;H^0_m)=\{z\in\mathbb{C}: |\re z|\le
|m|\Lambda\}$}.\hfill$\Diamond$
\end{REMARK}

\section{Proofs}\label{prfs}

Recall the statement of the Multiplicative Ergodic Theorem for the
cocycle $\{ D\varphi _t\}_{t\in \mathbb{R}}$, see \cite{Os}: For
each
 $\nu \in \mathcal{M}$ there exists a full
$\nu$-measure subset $X_\nu \subset \mathbb{T}^2$ such that for
each $x\in X_\nu$ and each nonzero $v \in
\mathcal{T}_x\mathbb{T}^2$, the tangent space at $x\in
\mathbb{T}^2$, the following forward and backward exact Lyapunov
exponents exist and are equal:
$$\lambda (x,v):=\lim_{t \to +\infty}t^{-1}\log
|D\varphi_t(x)v|=\lim_{t\to -\infty}t^{-1}\log |D\varphi
_t(x)v|.$$ Since the cocycle $\{D\varphi_t\}$ is two dimensional,
for each $x\in X_\nu$ and all $v\in \mathcal{T}_x\mathbb{T}^2$
there exist at most two different Lyapunov exponents that we will
denote by $\lambda_1$ and $\lambda_2$, $\lambda_1\geq \lambda_2$.
We stress that $\lambda_{1,2}=\lambda_{1,2}(\nu)$, and we denote
$\Lambda=\max\{\lambda_1(\nu):\nu\in\mathcal{M}\}$.

Since $\dv u=0$, we have $\det D\varphi_t(x)=1$ for all $x\in
\mathbb{T}^2$. This implies $\lambda_1+\lambda_2=0$ for all $x\in
X_\nu$ and $\nu\in \mathcal{M}$ \cite{Os}. If $y$ is a stagnation
point for $u$, then $D\varphi _t(y)=e^{tDu(y)}$ and $\re \sigma
(Du(y))=\{\lambda_1,\lambda_2\}$ for the Lyapunov exponents
$\lambda_{1,2}$ at $y$. Since $Du(y)$ is a matrix with real
entries and zero trace, if $\lambda_1\neq 0$ then, in fact,
$\sigma (Du(y))=\{ -\lambda _1,\lambda_1\}$.

\begin{REMARK}\label{MET1} {\em If $\lambda\in \Sigma\backslash \{0\}$ then
there exists a stagnation point $y$ such that $\lambda$ is a
Lyapunov exponent at $y$, cf. \cite{FSV99}. To see this, fix $\nu
\in \mathcal{M}$ and $x\in X_\nu$ such that $\lambda=\lambda
(x,v)$ for some $v\in \mathcal{T}_x\mathbb{T}^2\backslash\{0\}$.
Suppose that $u(x)\neq 0$. Since $\max_{x\in \mathbb{T}^2}|u(x)|<
\infty$, the identity
\begin{equation}\label{DuID} D\varphi_t(x)u(x)=u(\varphi_tx),
\quad t\in \mathbb{R}, \quad x\in\mathbb{T}^2,\end{equation}
implies that the forward Lyapunov exponent for $u(x)$ is
nonpositive. By the same reason the backward Lyapunov exponent for
$u(x)$ is nonnegative. Thus, $\lambda (x,u(x))=0$. This implies
$\lambda_1=\lambda_2=0$, in contradiction with $\lambda =\lambda
(x,v)\neq0$. Thus $x=y$, a stagnation point.}
\hfill$\Diamond$\end{REMARK}

\begin{REMARK}\label{MET2} {\em Assume that $\lambda
\in \Sigma $ and $\lambda >0$. By Remark~\ref{MET1}, find a
hyperbolic stagnation point $y$ such that
$\sigma(Du(y))=\{-\lambda ,\lambda\}$. If $v\in
\mathcal{T}_y\mathbb{T}^2$, $|v|=1$, is the eigenvector for
$Du(y)$ such that $Du(y)=-\lambda v$, then by the Stable Manifold
Theorem there is a manifold $\mathcal{O}$ that is tangent to $v$ at
$y$. In other words, if $x_0\to y$ such that $x_0\in \mathcal{O}$,
then $u(x_0)/|u(x_0)|\to v$. Also, for each $t\in \mathbb{R}$, if
$x_0\to y$, $x_0\in \mathcal{O}$, then $D\varphi_t(x_0)\to
e^{tDu(y)}$. Using \eqref{DuID}, we conclude that
\begin{equation}\label{Limu} \lim_{\mathcal{O}\ni x_0\to y}
u(\varphi_tx_0)/|u(x_0)|=e^{-\lambda t}v,\quad t\in
\mathbb{R}.\end{equation}} \hfill$\Diamond$\end{REMARK}

\begin{proof}[Proofs of Theorems~\ref{T:main} and \ref{T:main2}]
As explained in Section~\ref{mres}, it suffices to construct a weakly null
approximate eigenfunction $\{g\}$ for $A$ such that all functions
$g$ are localized along a streamline. For technical reasons it is
more convenient to work on the plane with straightened
streamlines. Therefore,  we introduce the necessary
volume-preserving change of variables localized around an orbit of
$\{\varphi_t\}$. After this change of variables $A$  becomes
simply the differentiation.

Fix an $N\in\mathbb{N}$ and a point $x_0\in \T^2$ such that
$u(x_0) \neq 0$ and $p(x_0)>3N$. For $u=(u_1, u_2)^\top$ denote
$u^\perp = ( -u_2 , u_1)^\top$.
 Let $\{\psi_\t\}$ be the local flow at $x_0$
such that
$$
\partial_\tau({\psi}_\t(x_0)) = \frac{u^\perp \circ \psi_\t(x_0)}{|u
\circ \psi_\t(x_0)|^2}.
$$
Define a mapping $H(t,\t) = \f_t \circ \psi_\t(x_0)$ for
$|t|<p(x_0)/2$ and $|\t|$ small enough to ensure the injectivity
of $H$. So, $H$ is defined on a horizontal strip
$\mathcal{S}=[-N,N]\times [-s,s]$. From the definition of $H$ we
obtain:
\def \p {\psi_\t(x_0)}
$$
DH(t,\t) = D \f_t (\p)\left[ \,\begin{matrix} u & \vdots &
\frac{u^\perp}{|u|^2} \end{matrix} \,\right] \circ \p.
$$
Thus, $\det{DH} = 1$ and $H$ is volume-preserving. Using this, one
easily computes %
\begin{multline}\label{E:H-Jac}
DH^{-\top}(t,\t) \overset{\text{def}}{=}DH^{-\top}(H(t,\t)) \\=
\kc{\p}\left[ \,\begin{matrix} \frac{u}{|u|^2}  & \vdots & u^\perp
\end{matrix} \,\right] \circ \p.
\end{multline}
\def \S {\mathcal{S}}
Given a function $F$ supported on $\S$ put $f = F\circ H^{-1}$. Then %
\begin{equation}\label{E:grad-1}
\n f(H(t,\t)) = DH^{-\top}(t,\t) \n F(t,\t).%
\end{equation}
More generally, for $D^mf\in\mathcal{B}_m$, the $m$-th differential
of $f$, by the chain rule (see, e. g., \cite[p.97]{AMR}) we have:
\begin{multline}\label{E:grad-m}
D^mf(H(t,\t))(v_1,\ldots,v_m) =
D^mF(t,\t)(DH^{-1}v_1,\ldots,DH^{-1}v_m) \\
+ \mbox{ lower order derivatives of }F.%
\end{multline}

Fix an $\a=\lambda+i\xi \in \C$. Define
\begin{equation}\label{AEFor}
F(t,\t) = e^{\a t}\g(t)\beta(\t)\quad\text{for}\quad
(t,\tau)\in[-N,N]\times[-s,s].
\end{equation}
A direct calculation shows:
\begin{equation}\label{A-action}
\rest{Af-\a f}{H(t,\t)} =\tilde{F}(t,\t),\quad\text{where}\quad
 \tilde{F}(t,\t)\overset{\text{def}}{=}e^{\a
t}\g'(t)\beta(\t).
\end{equation}
To make the main idea of the proof more transparent, we first
consider the case $m=1$.
 The general case $m\ge 1$ will be considered later.

We compute:
\begin{align}
\n F &= \left[ \begin{matrix} \a F + e^{\a t}\g'(t)\beta(\t) \\ e^{\a
t}\g(t)\beta'(\t)\end{matrix}\right] ;& \n \tilde{F}=\left[
\begin{matrix} \a \tilde{F} + e^{\a t}\g''(t)\beta(\t) \\ e^{\a
t}\g'(t)\beta'(\t)\end{matrix}\right].
\end{align}
Let us choose $\g(t) = (1-|t|N^{-1})\chi_{[-N,N]}$ smoothed out at
$\pm\, N,0$, and $\beta(\t) = (s - |\t|)\chi_{[-s,s]}$. Clearly,
$\frac{1}{2s}(\b'(\t))^2$ is an approximative kernel. So, if $s
\ra 0$, then
\begin{equation}\label{F-asym}
\begin{split}
\frac{1}{2s}|\partial F/\partial t|^2 & \ra 0, \quad
\frac{1}{2s}|\partial \tilde{F}/\partial t|^2 \ra 0  \text{ in }
L_1,\\
\frac{1}{2s}|\partial F/\partial \t|^2 & \ra e^{2\l
t}|\g(t)|^2\d_0(\t)\\
\frac{1}{2s}|\partial \tilde{F}/\partial \t|^2 &\ra e^{2\l
t}|\g'(t)|^2\d_0(\t).
\end{split}
\end{equation}
Here $\d_0$ denotes the Dirac $\d$-function, and the last two limits
in \eqref{F-asym} are understood in the sense of distributions.
Furthermore, the
measure of the support of $f=F\circ H^{-1}$ tends to zero as $s\ra
0$. So, $f/\|f\|_{H_1}$ converges to zero weakly, and is norm
bounded as $s \ra 0$. Since $K$ is a compact operator, we
therefore conclude that $\|K(f/\|f\|_{H_1})\|_{H_1} \ra 0$.
Passing to the $(t,\t)$-coordinates in
integrals and using \eqref{E:H-Jac} we have, as $s \ra 0$,
\begin{align}
\|L+\a\|_\bullet^2& \overset{\text{def}}{=} \inf\{\|Lg+\a g\|_{H_1}^2 :
\|g\|_{H_1} = 1\} \nonumber\\
&\lesssim \|Af - \a f\|_{H_1}^2 /\|f\|_{H_1}^2 + 
\|K(f/\|f\|_{H_1})\|_{H_1}^2 \nonumber\\
\intertext{by \eqref{E:grad-1} and \eqref{A-action},}%
 & = \frac{(2s)^{-1} \int_\mathcal{S} |DH^{-\top} \n
\tilde{F}|^2 d\t dt }{(2s)^{-1} \int_\mathcal{S} |DH^{-\top} \n
F|^2 d\t dt } +
\|K(f/\|f\|_{H_1})\|_{H_1}^2  \label{12_15}\\
\intertext{by \eqref{E:H-Jac} and \eqref{F-asym},}%
& \ra \frac{\int_\R |\kc{x_0}
u^{\perp}(x_0)|^2 e^{2\l t}|\g'(t)|^2 dt}{\int_\R |\kc{x_0}
u^{\perp}(x_0)|^2 e^{2\l t}|\g(t)|^2 dt}.\nonumber
\end{align}
Using the identity
$$\kc{x_0} u^{\perp}(x_0) = u^{\perp}\circ
\f_t(x_0),$$
the last expression is equal to
\begin{equation}\label{LE}
\frac{\int_\R |u\circ \f_t(x_0)|^2/|u(x_0)|^2 e^{2\l
t}|\g'(t)|^2 dt}{\int_\R |u\circ \f_t(x_0)|^2/|u(x_0)|^2 e^{2\l
t}|\g(t)|^2 dt}.
\end{equation}

Fix a nonzero Lyapunov exponent $\lambda\in\Sigma$ and any
$\xi\in\R$. By Remark \ref{MET1} there is a hyperbolic stagnation
point $y$ such that $\lambda$ is a Lyapunov exponent at $y$. Pick
$v\in\mathcal{T}_y\mathbb{T}^2$, $|v|=1$, such that
$D\varphi_t(y)v=e^{-\lambda t}v$. Assume for the moment that $\l
>0$. Using Remark \ref{MET2}, pick a nonperiodic point
$x_0$ that belongs to the manifold $\mathcal{O}$ tangent to $v$ at $y$. By
\eqref{Limu}, we have that $u\circ \f_t(x_0)/|u(x_0)|$ converges
to $e^{-\l t}v$ as $x_0 \ra y$ along this manifold. Use the
calculation above with $\alpha=\lambda+i\xi$. Passing to the limit
as $x_0\to y$, $x_0\in\mathcal{O}$, in \eqref{LE}, we obtain
\[
\|L+\a\|_\bullet^2 \leq \frac{\int_\R |\g'(t)|^2 dt}{\int_\R
|\g(t)|^2 dt}
\]
for arbitrary $N>0$. Observe that the quantity on the right hand
side tends to zero as $N\ra \infty$. The argument for $\l < 0$ is
similar.

Finally, to make $f$ mean-zero define another function $\bar{f}$
in the same way around same streamline and disjoint from $f$.
Varying its support we can obtain the equality $\int_{\T^2} f dx =
\int_{\T^2} \bar{f} dx$. Then $f-\bar{f}$ form the required
sequence of approximate eigenfunctions. In the sequel, we refer to
this procedure as symmetrization.

This finishes the proof of Theorem \ref{T:main} for $m=1$.

To prove part (a) in Theorem \ref{T:main2}, set $\a = i \xi$,
$\beta = \mathbf{1}_{[-s,s]}$ and keep $\g$ the same as before.
Then, as $s\ra 0$, we obtain:
\begin{align*}
\|L+\a\|_\bullet^2 &\lesssim \|Af - \a f\|_{L_2}^2 /\|f\|_{L_2}^2 +
\|K(f/\|f\|_{L_2})\|_{L_2}^2 \\
& = \frac{(2s)^{-1} \int_\mathcal{S} |\tilde{F}|^2 d\t dt
}{(2s)^{-1} \int_\mathcal{S} |F|^2 d\t dt } +
\|K(f/\|f\|_{L_2})\|_{L_2}^2 \ra  \frac{\int_\R |\g'(t)|^2
dt}{\int_\R |\g(t)|^2 dt}.
\end{align*}
Under the assumption on $\{\f_t\}$, $N$ can be taken arbitrarily
large. Symmetrization is carried out similarly. So, we have proved
that $i\R \ss \s_{\text{ess}}(L) \cap \s(A)$. On the other hand, since $A$ is
antisymmetric on $L_2^0$, $\s(A) \ss i \R$. Applying a version of
Weyl's Theorem as in \cite[Corollary XIII.4.2]{RS-IV}, we have
$\s_{\ess}(L)=\s_{\ess}(A) = i \R$.

We continue the proof of Theorem \ref{T:main}
 for $m\ge 1$.
Define $F$ as in \eqref{AEFor}, with
the same $\g$, $\a = m\l + i \xi$ and the cut-off function $\beta$
chosen such that if $s\ra 0$ then the following three conditions
are satisfied (see Appendix for a
construction of $\beta$):
\begin{itemize}
 \item[a)] for all
 $k=0,1,\ldots,m-1$ the derivatives $\beta^{(k)}(\t)$ tend to zero uniformly for
 $\t\in[-s,s]$;
 \item[b)] $|\rest{\beta^{(m)}}{[-s c,s c]}| > 1/2 $ for some fixed $c>0$;
 \item[c)] the norms $\| \beta^{(m)}\|_\infty$ are uniformly bounded.
 \end{itemize}
 This implies that, whenever $k<m$,
 \begin{equation}\label{F-asym-m2}
 \frac{1}{s}\left|\frac{\partial^{k+l}{F}}{\partial^l{t}
 \partial^k{\t}}\right|^2 \ra 0, \qquad
\frac{1}{s}\left|\frac{\partial^{k+l}{\tilde{F}}}
 {\partial^l{t}\partial^k{\t}}\right|^2 \ra 0.
\end{equation}
in $L_1$, as $s \ra 0$.

 On the other hand, for some sequence
$s_j \ra 0$,
\begin{equation}\label{F-asym-m}
\begin{split}
 \frac{1}{s_j}\left|\frac{\partial^{m}{F}}{\partial^m{\t}}\right|^2
 &\ra {c}e^{2\l m
 t} |\g(t)|^2 \d_0(t),\\
\frac{1}{s_j}\left|\frac{\partial^{m}{\tilde{F}}}{\partial^m{\t}}\right|^2
 &\ra {c}e^{2\l m
 t} |\g'(t)|^2 \d_0(t),
\end{split}
\end{equation}
and $\|K(f/\|f\|_{H_m})\|_{H_m} \ra 0$, where $f\circ H = F$.

As before we want to estimate the quantity $\|L+\a\|_\bullet^2$
and prove that it is zero. To this end, we notice 
(cf. \eqref{E:grad-m} and \eqref{12_15}) that
\begin{align}\label{est}
\|L+\a\|_\bullet^2 &\lesssim \frac{s_j^{-1} \int_\mathcal{S} \|D^m
\tilde{F}(DH^{-1}\cdot, \ldots, DH^{-1}\cdot)\|_{\mathcal{B}_m}^2
+ \| \text{lower} D\tilde{F}\|^2 d\t dt}{s_j^{-1} \int_\mathcal{S}
\|D^m F(DH^{-1}\cdot, \ldots, DH^{-1}\cdot)\|_{\mathcal{B}_m}^2 -
\| \text{lower} DF\|^2
d\t dt} \notag\\
&+ \|K(f / \|f\|_{H_m})\|_{H_m}^2 .
\end{align}
Our observations in \eqref{F-asym-m2} and \eqref{F-asym-m}
indicate that the only non-vanishing (in the limit as $s_j\ra 0$)
term under the integrals is
the one containing $\partial^m F/\partial^m \t$. More precisely,
denoting $w_j= \left\langle DH^{-1}v_j,\mathbf{e}_2
\right\rangle$, where $\mathbf{e}_2=(0,1)^\top$, we can express
this term as the product
$$
\frac{\partial^{m}{F}}{\partial^m{\t}} \cdot w_1 \cdot \ldots
\cdot w_m.
$$
Using formula \eqref{E:H-Jac} and the identity
$$
\kc{\psi_{\t}(x_0)} u^{\perp}(\psi_{\t}(x_0)) =
u^{\perp}(H(t,\t)),
$$
we also see that
\begin{align*}
w_j & =\left\langle v_j,DH^{-\top}\mathbf{e}_2 \right\rangle \\
& = \left\langle v_j, \kc{\psi_{\t}(x_0)}
u^{\perp}(\psi_{\t}(x_0))\right\rangle \\
& = \left\langle v_j, u^{\perp}\circ H \right\rangle.
\end{align*}
In particular,
$$
\sup_{|v_j| = 1}\left|\frac{\partial^{m}{F}}{\partial^m{\t}}
\cdot w_1 \cdot \ldots \cdot w_m\right|  \leq
\left|\frac{\partial^{m}{F}}{\partial^m{\t}}\right| \cdot
|u^{\perp}\circ H|^m.
$$
Thus, passing to the limit as $j\ra \infty$ in \eqref{est}, we
obtain
\begin{equation}\label{A:lastexpr}
\|L+\a\|_\bullet^2 \lesssim \frac{\int_\R |u^\perp\circ
\f_t(x_0)|^{2m} e^{2m\l t}|\g'(t)|^2 dt}{\int_\R |u^\perp\circ
\f_t(x_0)|^{2m} e^{2m\l t}|\g(t)|^2 dt}.
\end{equation}
The rest of the proof goes as in the case $m=1$.

To show (b) in Theorem \ref{T:main2},
 assume for the moment that $\{\f_t\}$ has arbitrary long
{\em periodic} orbits. Take a large $N$ and find an orbit
$\mathcal{O}$ with a finite period greater than $2N$. Pick a point
$x_0 \in \mathcal{O}$ such that
$$
|u(x_0)| = \max_{x \in \mathcal{O}} |u(x)| = c.
$$
Then for all $|t| \leq 1$ we have $ |u \circ \f_t(x_0)| \geq
\|D\f_{-t}(x_0)\|^{-1}c \geq Mc$, where $M$ depends only on
$\{\f_t\}$. Continuing from \eqref{A:lastexpr} with arbitrary $\a
\in i\R$ we obtain
$$
\|L+\a \|_\bullet^2 \lesssim {{N}^{-1}\ c^{2m}}/({M^{2m}c^{2m})},
$$
which gives the desired result.

Suppose now that for some $N > 0$ if $p(x)>N$ then $p(x)=\infty$
for the prime period $p(\cdot)$. Consider the set $S= p^{-1}((0,
\infty))$. The set $S$ is open and $p(x) \leq N$ for all
 $x \in \overline{S}$. Since $\{\f_t\}$ has arbitrarily
long orbits, $S \neq \T^2$. So, if $S \neq \emptyset$, then
$\partial{S} \backslash S \neq \emptyset$. This, however, leads to
a contradiction, since if $x \in \partial{S} \backslash S$, then
on one hand $p(x) \leq N$, but on the other hand $p(x) = \infty$.
We conclude that $S$ is empty and, as a consequence,
$p(x)= \infty$ for all $x\in\mathbb{T}^2$. In particular, $u(x)
\neq 0$ for all $x\in\T^2$. This allows us to bound
\eqref{A:lastexpr} with $\lambda=0$ by the expression
$$
c^{2m}\frac{\int_\R |\g'(t)|^2 dt}{\int_\R |\g(t)|^2 dt}, \text{
where } c = \frac{\max|u(x)|}{\min|u(x)|}.
$$
As before, we infer $\|L+\a\|_\bullet = 0$.
\end{proof}

\begin{proof}[Proof of Theorem~\ref{FESOM}] Since $L=-A+K$,
and the operator $K$ is compact,  Nussbaum's formula \eqref{nusbform} for essential spectral
radius implies that $\omega_{\ess}(L)=\omega_{\ess}(-A)$. Passing to
the vector field $-u$, it is enough to prove that
\begin{equation}\label{E:vishik+aux}
\omega (A)=\omega_{\ess}(A)=m\Lambda ,\quad m=1,2,\ldots,
\end{equation}
(recall, that $A$ is skew-self-adjoint on $L^0_2$, that is
$\omega(A)=\omega_{\ess}(A)=0$ for $m=0$).

Note that $\o_{\ess}(A)\ge 0$. Indeed, this follows from (b) in
Theorem \ref{T:main2}, provided $\{\f_t\}$ has arbitrary long
orbits. Otherwise, all orbits are periodic and the periods are
uniformly bounded. Excluding the trivial case $u \equiv 0$, pick a
point $x_0 \in \T^2$ such that $u(x_0) \neq 0$. Define a local
flow $\{\psi_\t\}$ as in the proof of Theorems
\ref{T:main}--\ref{T:main2} above. Take a smooth cut-off function
$\beta$ supported in a small interval $[-s,s]$. For
$x=\f_t(\psi_\t(x_0))$ set $f(x) = \beta(\t)$. After the
symmetrization  we obtain $f \in H_m^0$ and $e^{tA}f = f$. To see
that $1$ is an eigenvalue of $e^{tA}$ of {\em infinite}
multiplicity, we argue as follows. Take a periodic orbit. Make a
small transversal cross-section and split the cross-section into
infinitely many disjoint intervals. For each interval $I$
construct a function $f$ as above such that $e^{tA}f = f$ and the
support of $f$ belongs to the orbit of $I$. Thus,
$1\in\sigma_{\ess}(e^{tA})$ as claimed.

Therefore, in the proof of the inequality $\omega_{\ess}(A)\geq
m\Lambda$ we may assume that $\Lambda >0$. By Theorem~\ref{T:main}
for $A$ and the spectral inclusion $e^{t\sigma (A)}\subset \sigma
(e^{tA})$ we have that $e^{t(m\lambda+i\xi)}\subset \sigma
_{\ess}(e^{tA})$ for each nonzero $\lambda \in \Sigma$ and all
$\xi \in \mathbb{R}$. Thus $m\Lambda \leq \omega_{\ess}(A)$.
It remains to prove that $\omega (A)\leq m\Lambda$. This is
implied by the following lemma, proved in Appendix.
\begin{LEMMA}\label{LyapExGen}
\begin{equation}\label{LEsT}
m\Lambda\ge\lim_{t\to \infty}t^{-1}\log \max_{x\in \mathbb{T}^2}\|
D^m\varphi_t(x)\|_{\mathcal{B}_m}, \quad m=1,2,\ldots.
\end{equation}
\end{LEMMA}
Indeed, by \eqref{LEsT} for each $m=1,2,\ldots $ and each
$\epsilon >0$ there is a constant $c=c(\epsilon ,m)$ such that
\begin{equation}\label{AD2.1} \max_{x\in \mathbb{T}^2}
\| D^m(\varphi_k)(x)\|_{\mathcal{B}_m}\leq ce^{mk(\Lambda
+\epsilon)}\text{ for } k=1,2,\ldots .\end{equation}
 It suffices to prove that $\| f\circ \varphi_k\|_{H_m}\leq
ce^{m(\Lambda +\epsilon )k}\| f\|_{H_m}$ for all $k\in\mathbb{N}$.
Note that
$$\| f\circ \varphi_k\|^2_{H_m}=
\int_\mathbb{T}\max_{ 0\leq n\leq m} \| D^n(f\circ
\varphi_k)(x)\|^2_{\mathcal{B}_n}dx,$$ and apply to $f\circ
\varphi_k$ the chain rule in \cite[p. 97]{AMR}. Using
\eqref{AD2.1} for $m=j_q$ we have the desired result:
\begin{equation*}\begin{split} &\|D^n (f\circ \varphi_k)(x)\|
_{\mathcal{B}_n}\\
&\quad \leq \sum^n_{p=1}\sum_{j_1+\dotsb +j_p=n}\sum_{\{\ell\}} \|
D^pf(\varphi_kx)\|_{\mathcal{B}_p}\prod^p_{q=1}\|
D^{j_q}(\varphi_k)(x)\|_{\mathcal{B}_{j_q}}\\
&\quad\leq \sum^n_{p=1}\sum_{j_1+\dotsb+j_p=n}\sum_{\{\ell\}}\|
D^pf(\varphi_kx)\|_{\mathcal{B}_p}c\exp [(\Lambda+\epsilon
)(j_1+\dotsb
+j_p)k]\\
&\quad\leq c(\epsilon ,m)\max_{1\leq n\leq m}\|
D^nf(\varphi_kx)\|_{\mathcal{B}_n}\exp [(\Lambda +\epsilon
)mk].\end{split}\end{equation*} Here the summation $\sum_{\{e\}}$
for each $p$ is taken over $\ell_1<\dotsb < \ell_{j_1}$, $\dotsc$,
$\ell_{j_1+\dotsb +j_{p-1}+1}<\dotsb <\ell_p$, see \cite[p.
97]{AMR}.\end{proof}

\section*{Appendixes}

\begin{proof}[{\bf 1.} The construction of $\beta$]
Select $\phi\in C^\infty_0[-1,1]$ with
$\rest{\phi}{[-1/2,1/2]}\equiv 1$.
Let $C=\|\phi\|_{C^m}$. We define $ \beta_1(\t) = \phi (\t/s)$ and
$\beta_2(\t)=\t^m / m!$, and set $\beta=\beta_1\beta_2$. Since
$\|\beta_1^{(l)}\|_{\infty} \leq C/s^l$, we have $\beta^{(k)} =
\sum_{l=0}^k C_k^l \beta_1^{(l)} \beta_2^{(k-l)}$ and
\[|\beta^{(k)}(\t)| \leq \sum_{l=0}^k
C_k^l\frac{1}{s^l}\frac{|\t|^{m-k+l}}{(m-k+l)!} \lesssim |\t|
^{m-k}.\] Thus, conditions a) and c) are fulfilled. Notice also
that
\[|\beta^{(m)}(\t)| \geq |\beta_1(\t)| - C\sum_{l=1}^m
C_{m}^{l}\left({|\t|}{s^{-1}}\right)^l/{l!}.\]
So, if $0 < c < 1/2$ is such that $C\sum_{l=1}^m C_{m}^{l}
{c^l}/{l!} < 1/2$, then $|\rest{\beta^m}{[-s c,s c]}|
> 1/2$, and b) is proved.\end{proof}

\begin{proof}[{\bf 2.} Proof of Proposition \ref{THM4.1}]
Let us assume the contrary. First, we rule out one simple case:
there is an unstable stagnation point $x_0$ in the sense of
Lyapunov, namely: {\em There exists a neighborhood $U$ of $x_0$
such that for some sequence $x_n \ra x_0$ and $t_n\in\mathbb{R}$
 we have $ \f_{t_n}(x_n)
\notin U$}. By our assumption the sequence $\{t_n\}$ is bounded.
For a limit point $t$ then
$\f_{t_n}(x_n) \ra \f_{t}(x_0) = x_0 \notin U$, which is a
contradiction.

Fix a point $x_0 \in \T^2$ such that $u(x_0) \neq 0$.  Let $U$
denote the open connected component in the set $\{x: u(x) \neq 0
\}$ containing $x_0$. Note that $U$ is linearly connected.
Take a continuous path $\g : [0,1] \ra
\T^2$ such that $\g((0,1)) \ss U$,
 and $y_0=\g(0)$ and $y_1=\g(1)$
 belong to the boundary $\partial
U$,  $y_0\ne y_1$.
Choose neighborhoods $U_0$ and $U_1$ of $y_0$ and $y_1$,
respectively, such that $\g ([1/3,2/3]) \ss \T^2 \backslash (U_0
\cup U_1)$. By our assumption, $y_0$ and $y_1$ are Lyapunov
stable. This implies that there are two orbits $\mathcal{O}_0 \ss
U_0$ and $\mathcal{O}_1 \ss U_1$ intersecting $\g$. Find the
smallest $s_0$ (largest $s_1$) in $[0,1]$, for which $\g(s_0) \in
\mathcal{O}_0$ ($\g(s_1) \in \mathcal{O}_1$). Then $s_0 < s_1$ and
$\g((s_0,s_1))$ lies in the exteriors of the closed curves
$\mathcal{O}_0,\mathcal{O}_1$.

The prime period function $p(\cdot)$ is continuous on the set $\{x
\in \T^2: p(x)>0\}$. Using this fact we define a continuous
function $h: [s_0,s_1] \times [0,1] \ra \T^2$ such that $h(s,t) =
\f_{t p(\g(s))}(\g(s))$. Since $ \cup_{0\leq t\leq 1}h(s_j,t) =
\mathcal{O}_j$, $j=1,2$, we conclude that $h$ defines a continuous
homotopy between $\mathcal{O}_0$ and $\mathcal{O}_1$. We claim
that the image of $h$ does not intersect the interiors of
$\mathcal{O}_0$ and $\mathcal{O}_1$.
To prove the claim, suppose that $h(s',t')$ belongs, say, to the
interior of $\mathcal{O}_0$. Then clearly $s' \in (s_0,s_1)$. By
our construction this means that $\g$ lies in the exterior of
$\mathcal{O}_0$. So, $\g(s') \in \text{exterior }\mathcal{O}_0$
and $\f_{t' p(\g(s'))}(\g(s')) \in \text{interior }\mathcal{O}_0$.
This implies the existence of a point $t'' \in[0,1]$ such that
$\f_{t'' p(\g(s'))}(\g(s')) \in \mathcal{O}_0$ and hence, $\g(s')
\in \mathcal{O}_0$, a contradiction. This proves the claim.

To finish the proof of the proposition, attach another torus to
$\mathbb{T}^2$ along the curve $\mathcal{O}_0$. We obtain a
double-torus $\T^2 + \T^2$. It follows from the claim above that
$h$ is a homotopy of $\mathcal{O}_0$ into $\mathcal{O}_1$
on $\T^2 + \T^2$. This is not possible, because the loops
$\mathcal{O}_0$ and $\mathcal{O}_1$ belong to different classes of
the fundamental group on $\T^2 + \T^2$. This contradiction
finishes the proof. Note that the same result with the identical
proof holds on any domain in $\R^2$.\end{proof}

\begin{proof}[{\bf 3.} Proof of Lemma \ref{LyapExGen}]
We prove \eqref{LEsT} by induction and take $t=k\in\mathbb{N}$ in
\eqref{LEsT}. For $m=1$ this inequality (and even equality) is
given in \eqref{LyapF}, see \cite[Thm. 8.15]{CL}. Assume that
\eqref{LEsT} holds for $m=1,\dotsc ,n-1$. For $k\in\mathbb{N}$ use
the chain rule in \cite[p. 97]{AMR} for $\varphi \circ
\varphi_{k-1}$:
\begin{align}
& D^n(\varphi \circ \varphi_{k-1})(x)(v_1,\dotsc ,v_n)\notag\\
& \quad =D^n\varphi (\varphi_{k-1}x)(D(\varphi_{k-1})(x)v_1,
\ldots,D(\varphi_{k-1})(x)v_n)\notag\\
&\qquad\qquad + \sum^{n-1}_{p=2}\sum_{j_1+\dotsc
+j_p=n}\sum_{\{\ell\}}
D^p\varphi (\varphi_{k-1}x)\label{AD2.2}\\
& \Big(D^{j_1}(\varphi_{k-1})(x)(v_{\ell_1},\dotsc
,v_{\ell_{j_1}}),\dotsc ,
 D^{j_p}(\varphi_{k-1})(x) (v_{\ell_{j_1+\dotsb +j_{p-1}+1}} ,
\dotsc ,v_{\ell_p})\Big)\notag\\
&\quad + D\varphi (\varphi_{k-1}x)D^n(\varphi_{k-1})(x)(v_1,\dotsc
,v_n).\notag
\end{align}
The middle term in \eqref{AD2.2} does not contain derivatives of
$\varphi$ of order $n$. By the induction assumption, we may apply
estimate \eqref{AD2.1} for $m=1,\dotsc ,n-1$. Since the number of
summands in the middle term does not depend on $k$, we conclude
that the norm of all terms in \eqref{AD2.2} except the last one is
dominated by $c\exp [n(\Lambda +\epsilon )k]$. For the last term
we again use the chain rule for $\varphi_{k-1}=\varphi\circ
\varphi_{k-2}$:
\begin{equation*}\begin{split} & D\varphi (\varphi_{k-1}x)D^n(\varphi
\circ \varphi_{k-2})(x)
(v_1,\dotsc ,v_n)\\
&\quad =D\varphi (\varphi_{k-1}x)D^n\varphi
(\varphi_{k-2}x)(D(\varphi_{k-2})(x)v_1,\dotsc ,
D(\varphi_{k-2})(x)v_n)\\
&\qquad + \sum^{n-1}_{p=2}\sum_{j_1+\dotsb + j_p=n}\sum_{\{\ell\}}
D\varphi (\varphi _{k-1}x)D^p\varphi
(\varphi_{k-2}x)\\
&\qquad\qquad\Big(D^{j_1}(\varphi_{k-2})(x)(\ldots ),\dotsc ,
D^{j_p}(\varphi _{k-2})(x)(\ldots)\Big)\\
&\qquad +D\varphi (\varphi_{k-1}x)D\varphi
(\varphi_{k-2}x)D^n(\varphi_{k-2})(x) (v_1,\dotsc
,v_n).\end{split}\end{equation*}

Again, by the induction assumption and \eqref{AD2.1}, the norms of
all terms, except the last one, are dominated by $c\exp [n(\Lambda
+\epsilon)k]$. But the last term can be written as
$D(\varphi_2)(\varphi_{k-2}x)D^n(\varphi_{k-2})(x)(\ldots )$. We
repeat this argument until the last term becomes
$D(\varphi_{k-1})(\varphi x)D^n\varphi (x)(\ldots )$. Since the
total number of terms is growing
 not faster than polynomially in $k$,
 \eqref{LEsT} follows.\end{proof}

\end{document}